\documentclass[preprint,authoryear,12pt]{elsarticle}
\usepackage{graphicx}
\usepackage{a4wide}
\usepackage{epstopdf}
\usepackage{booktabs}
\usepackage{natbib} 
\usepackage{multirow}
\usepackage{amssymb}
 \usepackage{amsthm}
\usepackage{color,soul} 
\definecolor{lightred}{rgb}{1,.6,.6}
\sethlcolor{green}
\setulcolor{red}
\setstcolor{red}

\begin{document}

\begin{frontmatter}

\title{Time-Frequency Dynamics  of Biofuels-Fuels-Food System}

\author[utia,ies]{Lukas Vacha} \ead{vachal@utia.cas.cz}
\author[ies,vse,CERGE]{Karel Janda} \ead{Karel-Janda@seznam.cz}
\author[utia,ies]{Ladislav Kristoufek} \ead{kristouf@utia.cas.cz}
\author[berkeley]{David Zilberman} \ead{zilber11@berkeley.edu}

\address[utia]{Institute of Information Theory and Automation, Academy of Sciences of the Czech Republic, Pod Vodarenskou Vezi 4, 182 08, Prague, Czech Republic, EU} 
\address[ies]{Institute of Economic Studies, Faculty of Social Sciences, Charles University in Prague, Opletalova 26, 110 00, Prague, Czech Republic, EU}
\address[vse]{Department of Banking and Insurance, Faculty of Finance and Accounting, University of Economics, Prague, Namesti Winstona Churchilla 4, 130 67, Prague, Czech Republic, EU}
\address[CERGE]{Center for Economic Research and Graduate Education, joint workplace of Charles University and the Academy of Sciences of the Czech Republic, Politickych veznu 7, 111 21, Prague, Czech Republic, EU}
\address[berkeley]{Department of Agricultural and Resource Economics, University of California, Berkeley, 207 Giannini Hall, Berkeley, California  94720, USA}

\begin{abstract}

For the first time, we apply the wavelet coherence methodology on biofuels (ethanol and biodiesel) and a wide range of related commodities (gasoline, diesel, crude oil, corn, wheat, soybeans, sugarcane and rapeseed oil). This way, we are able to investigate dynamics of correlations in time and across scales (frequencies) with a model-free approach. We show that correlations indeed vary in time and across frequencies. We find two highly correlated pairs which are strongly connected at low frequencies -- ethanol with corn and biodiesel with German diesel -- during almost the whole analyzed period (2003-2011). Structure of correlations remarkably changes during the food crisis -- higher frequencies become important for both mentioned pairs. This implies that during stable periods, ethanol is correlated with corn and biodiesel is correlated with German diesel mainly at low frequencies so that they follow a common long-term trend. However, in the crisis periods, ethanol (biodiesel) is lead by corn (German diesel) even at high frequencies (low scales), which implies that the biofuels prices react more rapidly to the changes in their producing factors.

\end{abstract}

\begin{keyword}
biofuels \sep prices \sep correlations \sep wavelet coherence\\
\textit{JEL codes:} C22, Q16, Q42
\end{keyword}

\end{frontmatter}

\newpage
\section{Introduction}

Relationship between biofuels and related fossil fuels and producing agricultural commodities and its analysis have become more challenging to study in recent years which experienced strongly varying prices of all mentioned commodities. The so-called ``food crisis", which was characteristic by sharply increasing prices of agricultural commodities and crude oil as well as retail fuels and biofuels, captured a very wide academic and policy attention during 2008 and it continues to form policy attitude toward the biofuels versus food issues. The matter of food--fuels--biofuels interactions gained another dimension and a research on possible squeeze-out effect, i.e. whether the increasing prices of biofuels cause prices of related agricultural commodities to raise as well, has become very frequent since that time. 

However, the results are in general quite ambiguous, which might be caused by the fact that the authors usually use different models with different assumptions and restricted commodity coverage coming to different results \citep{Janda_Kristoufek_and_Zilberman_2012,Zilberman_Hochman_Rajagopal_Sexton_and_Timilsina_2012}. In this paper, we contribute to this discussion by providing a new comprehensive view on the price-correlation dynamics of food-biofuels-fuels system. Using the wavelet coherence analysis, we are able to capture complex price-correlation dynamics without restriction to ad-hoc specified time or frequency frameworks used in the previous literature. Additional advantage of our paper is a wide coverage of all biofuels related commodities including crude oil, fossil fuels, both main types of biofuels and major agricultural feedstocks for biofuels, which is a unique contribution to biofuels price transmission literature.

In the previous studies, \cite{Zhang_Lohr_Escalante_and_Wetzstein_2009,Zhang_Lohr_Escalante_and_Wetzstein_2010} use VECM and mGARCH models to analyze the US ethanol connections with corn, soybeans and gasoline to find no long-range relationships. Also, they focus on Granger causality and uncover only weak short-term effects. 
\cite{McPhail_2011} uses structural VAR model to analyze relationship between the US ethanol, crude oil and gasoline to show that a policy-driven increase in demand for ethanol leads to lowering prices of both crude oil and gasoline. \cite{Busse_Brummer_and_Ihle_2010} focus on German biodiesel and its connections to rapeseed oil, soy oil and crude oil between 2002 and 2009 and argue that crude oil strongly influenced the prices of biodiesel and biodiesel shocks transmitted into rapeseed oil prices. However, the results are regime-dependent. 

A number of previous studies dealing with price transmission in food-energy systems do not consider the prices of biofuels at all. Instead they just consider crude oil prices and prices of agricultural commodities. \cite{Ciaian_and_Kancs_2011} report cointegration between crude oil and a range of food commodites, some of them being used in the production of biofuels. Since their range of food commodities cointegrated with the prices of crude oil grows over time, they provide supporting evidence to the hypothesis of increasing importance of biofuels transmission channel in the link between energy and food markets \citep{Tyner_2010,Ciaian_and_Kancs_2011_FP}.

\cite{Serra_Zilberman_Gil_and_Goodwin_2010, Serra_Zilberman_Gil_and_Goodwin_2011} and \cite{Serra_Zilberman_and_Gil_2011} focus on cointegration between crude oil, ethanol and related feedstock to find an equilibrium relationship between the commodities for the US market as well as the Brazilian market with a slower reaction to the shocks found for the latter. \cite{Rajcaniova_and_Pokrivcak_2011_EC} argue that the cointegration relationship between ethanol and related commodities exists only for years 2008 onwards, finding no statistically significant relationship in preceding years 2005--2008. \cite{Pokrivcak_and_Rajcaniova_2011} provide evidence for cointegration relationship between crude oil and gasoline prices but they do not find any cointegration between the prices of ethanol and gasoline, and ethanol and oil.

\cite{Kristoufek2012a} analyze the biofuels markets with a use of minimum spanning and hierarchical trees to show that biofuels are very weakly connected to fossil fuels and relevant agricultural commodities in short-term but become more interrelated in medium-term. The relations become stronger for the food crisis period onwards. \cite{Kristoufek2012} study the same dataset as the previous reference and focus on elasticities and Granger causality and their price dependence. They find that corn causes changes in ethanol prices while both elasticity and causality are price-dependent, and they find biodiesel to be caused and elastic to the changes in German diesel prices and the effects are again price-dependent.

Evidently, the results differ considerably not only due to the model specifics but also due to the analysis of sometimes different time scales (the most frequently analyzed scales range from weekly to monthly or quarterly). Moreover, standardly used time series econometric methods usually consider the frequency and time components separately. In this paper, we utilize the wavelet approach, which allows to study the frequency components of time series without losing the time information. Moreover, the wavelet methodology is constructed to work with nonstationary data, which is a frequent issue in the financial time series modeling \citep{Roueff2011}. 

We are the first ones to apply the wavelet coherence analysis on biofuels (ethanol and biodiesel) and a wide range of related commodities (gasoline, diesel, crude oil, corn, wheat, soybeans, sugarcane and rapeseed oil). Wavelets have been used several times in the analysis of commodities and energy markets. \cite{ConnorRossiter} were among the first ones to use wavelets on the commodity markets. They studied price correlations using a discrete form of wavelet transform. Relations between oil prices and economic activity with wavelets were studied by \cite{Naccache2011}. However, that study was focused on very long cycles. Recently, \cite{VachaBar2012} applied continuous wavelet analysis to study dynamic dependence between energy commodities. The wavelet method was compared with multivariate concept of dynamic conditional correlation generalized autoregressive conditional heteroscedasticity (DCC-GARCH).

We show that correlations indeed vary in time and across frequencies. We find two highly correlated pairs which are strongly connected during almost the whole analyzed period (2003-2011) at low frequencies -- ethanol with corn and biodiesel with German diesel. This asymmetric behavior of ethanol and biodiesel is quite an interesting phenomenon since a simple technological reasoning could assume that both biofuels would have similar correlation structures with respect to their agricultural feedstock and appropriate fossil fuel substitute. However we show that this is not the case and that ethanol prices are primarily connected with the price of its major US feedstock while the biodiesel prices are most strongly connected with prices of its German fossil fuel substitute.

We discover that structure of correlations remarkably changes during the food crisis -- higher frequencies become important for both mentioned pairs. This implies that during the stable periods, ethanol is correlated with corn and biodiesel is correlated with German diesel mainly at low frequencies so that they follow a common long-term trend. However, in the crisis periods, ethanol (biodiesel) is lead by corn (German diesel) even at high frequencies (low scales), which implies that the biofuels prices react more rapidly to changes in their producing factors.

The rest of this paper is structured as follows. In Section 2, we provide the basic definitions of the wavelet analysis -- wavelets, wavelet transforms and coherence. In Section 3, the analyzed dataset is described and the results of wavelet analysis are discussed. Section 4 concludes.

\section{Methodology} 

In this section, we briefly introduce the continuous wavelet transform, wavelet coherence and wavelet phase differences. The wavelet transform decomposes the time series from a time-domain to a time-frequency domain, i.e. using wavelets, we transform one dimensional time series into a two-dimensional space. Contrary to the Fourier transform, the wavelet transform uses a localized function with finite support -- a wavelet -- for the decomposition. For this reason, wavelet transform has significant advantages over the Fourier transform mainly when the object under study is nonstationary, or only locally stationary \citep{Roueff2011}. In the case we use just the Fourier transform, we obtain only the information about the frequency components, but we completely loose the time information. Therefore, in case a change in behavior arises in the middle of the investigated time series, we are not able to localize where exactly this change occurs. When bivariate relation is studied, the same problem with time localization applies, see \cite{Gencay2002, PercivalWalden2000,Ramsay2002,VachaBar2012} for details. The utilized wavelet analysis overcomes these issues. Since the biofuels markets are relatively new, their behavior is very dynamic and unstable as will be visible in the following sections and is as well documented in \cite{Kristoufek2012a,Kristoufek2012}. Thus the need for the localized time-frequency wavelet analysis of biofuels and related commodities is clearly evident.

\subsection{The continuous wavelet transform}

The continuous wavelet transform $W_x(u,s)$ is defined as a projection of a specific wavelet\footnote{We use the Morlet wavelet that belongs to the family of complex wavelets. Complex, or analytical, wavelets have a real and a complex part, hence we can perform the phase analysis. The Morlet wavelet is defined as $\psi^M(t)=\frac{1}{\pi^{1/4}}e^{i\omega_0 t}e^{-t^2/2}$, where $\omega_0$ denotes the central frequency of the wavelet.  In our analysis, we set $\omega_0=6$, which is the value often used in the economic applications \citep{RuaNunes2009}. } $\psi(.)\in L^2(\mathbb{R})$ onto the examined time series $x(t)\in L^2(\mathbb{R})$,
\begin{equation}
W_x (u,s)=\int_{-\infty}^\infty x(t)\frac{1}{\sqrt s}\overline{\psi \left( \frac{t-u}{s}\right)} dt,
\label{eq7}
\end{equation}
where $u$ determines the exact position of the wavelet\footnote{This parameter helps to perfectly localize the behavior of the time series under study. In other words, this is the extra parameter that we do not have in the Fourier transform.}.
Scale parameter $s$ controls how the wavelet is stretched or dilated. If the scale is lower (higher), the wavelet is more (less) compressed, therefore the wavelet is able to detect higher (lower) frequency components of the examined time series $x(t)$. A wavelet must fulfill the admissibility condition: $C_{\psi}=\int_{0}^{\infty }\frac{\mid\Psi(f)\mid^2}{f}df<\infty$, where $\Psi(f)$ is the Fourier transform of a wavelet $\psi(.)$. The time series $x(t)$ can be reconstructed using the wavelet coefficients as
\begin{equation}
x(t)=\frac{1}{C_\psi}\int_0^\infty \left[\int_{-\infty}^\infty  W_x (u,s) \psi_{u,s}(t) du \right] \frac{ds}{s^2} ,\hspace{5 mm}s>0.
\end{equation}
Importantly, the continuous wavelet transform preserves energy of the analyzed time series, i.e.,
\begin{equation}
\| x \|^2 =\frac{1}{C_\psi}\int_0^\infty  \left[\int_{-\infty}^\infty  \left| W_x (u,s)\right|^2 du \right] \frac{ds}{s^2}.
\end{equation}
We use this key property for the definition of the cross wavelet power and subsequently of the wavelet coherence. For a more detailed introduction to wavelets, see \cite{Daubechies1988, PercivalWalden2000}.

\begin{figure}[htbp]
\center
\begin{tabular}{c}
\includegraphics[width=5in]{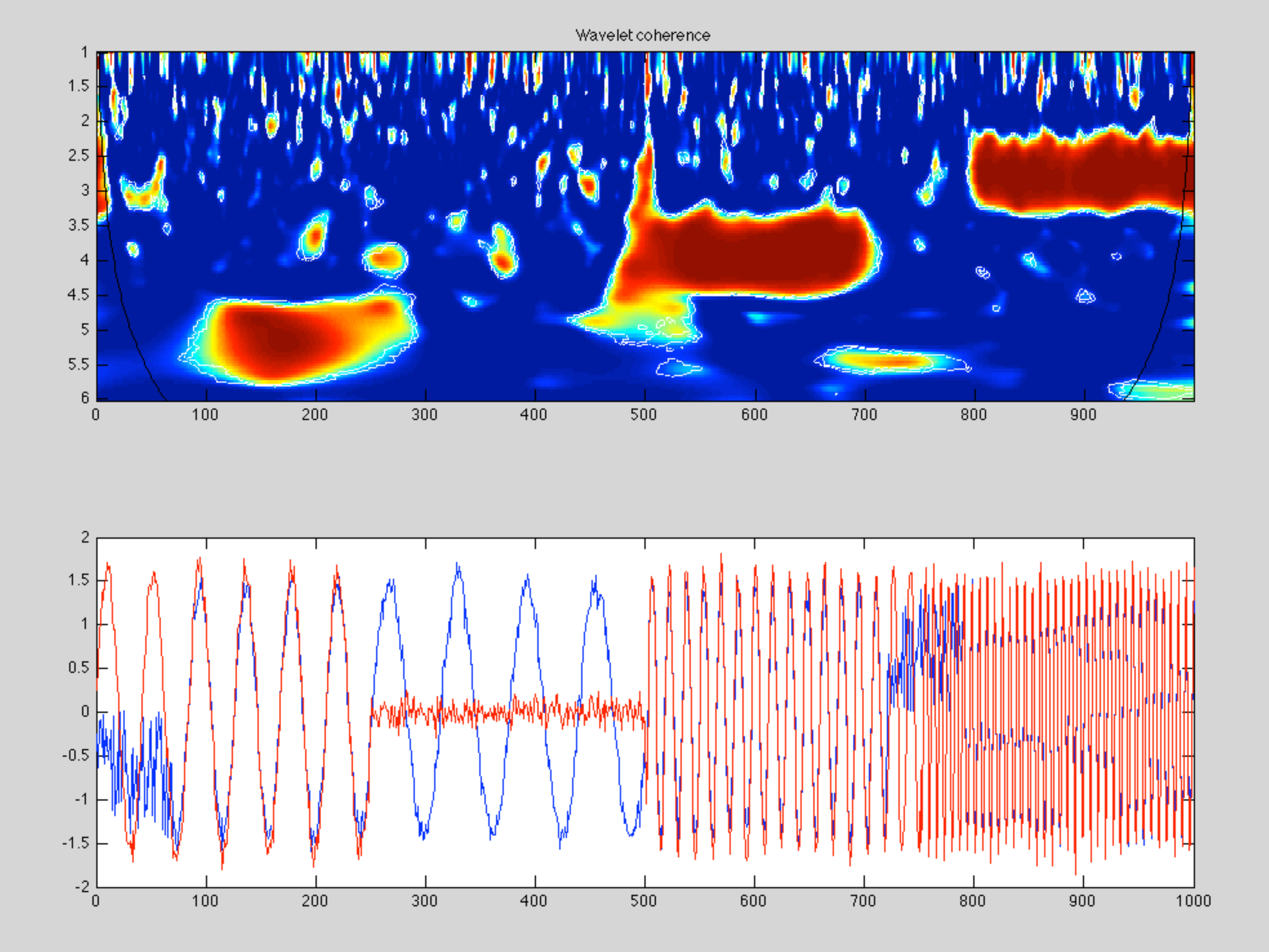}\\
\end{tabular}
\caption{\footnotesize\textit{Wavelet coherence example.}\label{excoher}}
\label{coh}
\end{figure}

\subsection{Wavelet coherence}

Since we study the interactions between two time series, we introduce a bivariate setting called wavelet coherence. As the first step, we define the cross wavelet transform and subsequently the cross wavelet power. 

The cross wavelet transform \citep{TorenceCompo98} of two time series $x(t)$ and $y(t)$ is defined as
\begin{equation}
W_{xy} (u,s) = W_x (u,s) \overline{W_y (u,s)},
\end{equation}
where $W_x (u,s)$ and $W_y (u,s)$ denote the continuous wavelet transforms of $x(t)$ and $y(t)$, respectively, $u$ defines a time position, and $s$ denotes the scale parameter. Further, using the cross wavelet transform, we obtain the cross wavelet power as $|W_{xy} (u,s)|$.
The cross wavelet power represents the local covariance between the examined time series at the specific scale $u$. In other words, it indicates where the time series have high common power in the time-frequency domain.

Following \cite{TorrenceWebster99}, we define the squared wavelet coherence coefficient as: 
\begin{equation}
R^2 (u,s)=\frac{|S(s^{-1}W_{xy} (u,s))|^2}{S(s^{-1}|W_x (u,s)|^2) S(s^{-1}|W_y (u,s)|^2)},
\end{equation}
where $S$ is a smoothing operator\footnote{Smoothing is achieved by convolution in both time and scale, see \cite{Grinsted2004} for more details.}.
The squared wavelet coherence coefficient is in the range $0\le R^2 (u,s) \le1$. Values of the coherence close to one indicate strong correlation at a given scale, while values close to zero indicate no correlation. The squared wavelet coherence can be perceived as a local linear correlation between two time series at a particular scale. Fig. \ref{excoher} shows example of the wavelet coherence on three different scales and at different time positions. 

We test statistical significance of the wavelet coherence estimates using Monte Carlo methods. The testing procedure is based on the approach of \cite{Grinsted2004} and \cite{TorenceCompo98}. The significant areas of the wavelet coherence are bordered with black thick line.

Since wavelets are in fact filters, we have to deal with boundary conditions. This problem arises at the beginning and at the end of a dataset, where the filter analyzes nonexistent data. In our work, we solve this problem by augmenting the dataset with a sufficient number of zeros. The area where we pad the dataset with zeros is called the cone of influence.  It is graphically represented by a cone bordered by a bold black line in our figures. For more details, see \cite{TorenceCompo98}, \cite{Grinsted2004}.

\begin{figure}[htbp]
\center
\begin{tabular}{c}
\includegraphics[width=5in]{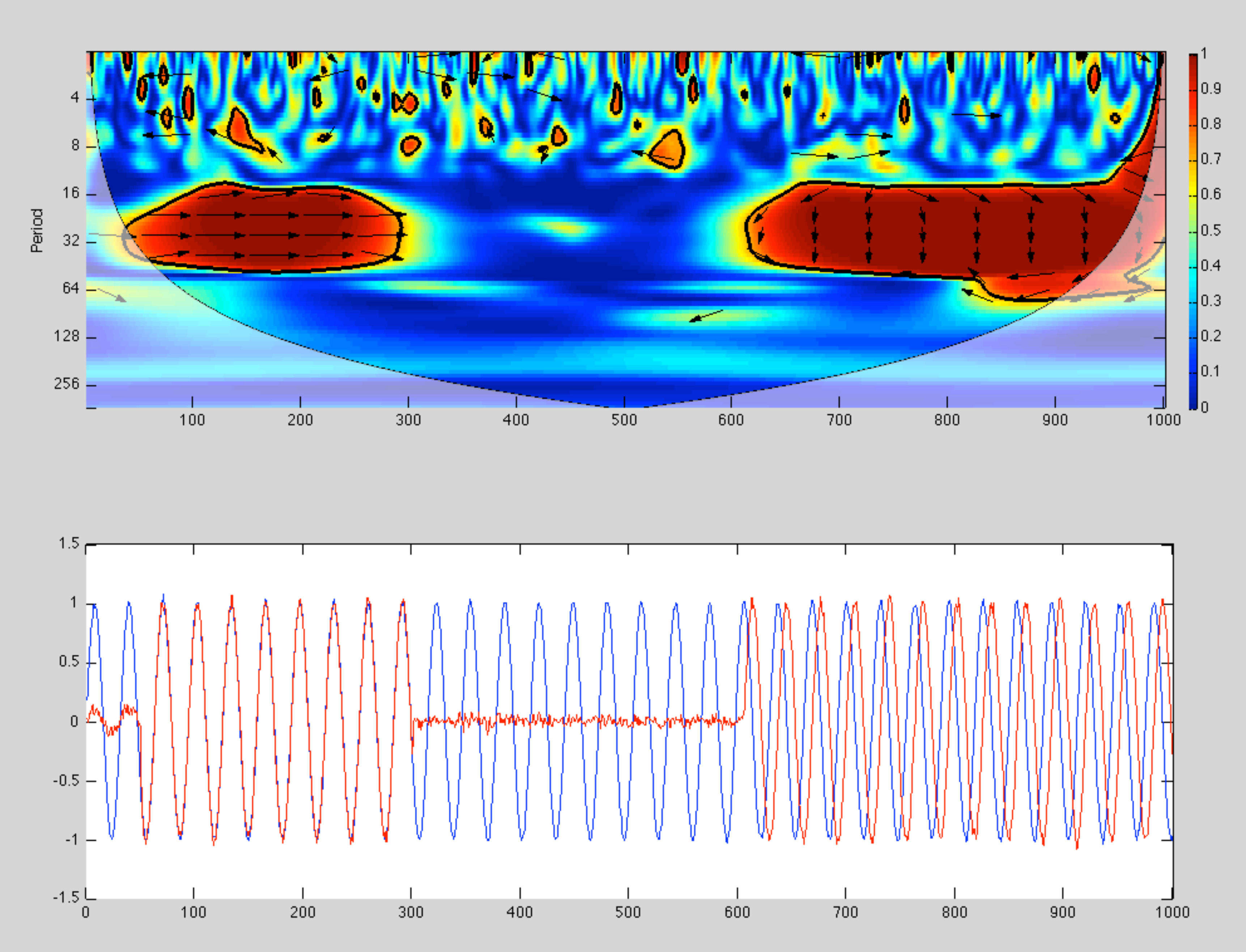}\\
\end{tabular}
\caption{\footnotesize\textit{Wavelet phase example.}\label{exphase}}
\label{phase}
\end{figure}

\subsection{Phase}

Since the wavelet coherence coefficient is squared, we cannot distinguish between negative and positive correlation. For this reason, we use the wavelet coherence phase differences which indicate delays in the oscillation between the two examined time series. Following \cite{TorrenceWebster99}, we define the wavelet coherence phase difference as
\begin{equation}
\phi_{xy} (u,s)=\tan ^{-1}\left( \frac{\Im \{S(s^{-1}W_{xy} (u,s))\}}{\Re \{S(s^{-1}W_{xy} (u,s))\} }\right).
\end{equation}
Phase differences are indicated by black arrows in our figures. In case the two examined time series move together, they have a zero phase difference on a particular scale and the arrows point to the right. If the time series are in anti-phase, i.e, they are negatively correlated, then the arrows point to the left. Arrows pointing down indicate that the first time series leads the second one by
$\frac{\pi}{2}$, 
whereas arrows pointing up means that the second time series leads the first one by 
$\frac{\pi}{2}$. 
A mixture of positions is common.  For example, an arrow pointing up and right means that the time series are in phase, with the second time series leading the first one. As an illustration, see Fig. \ref{exphase} where the case of zero phase difference and phase shift by $\frac{\pi}{2}$ 
are depicted.

\section{Data and results}

\begin{table}[htbp]
\centering
\caption{Analyzed commodities}
\label{tab1}
\footnotesize
\begin{tabular}{c|c|c}
\toprule \toprule
Commodity&Ticker&Contract type\\
\midrule \midrule
Biodiesel & BIOCEUGE & Spot, Germany \\
Corn& C1 & Futures, CBOT\\
Crude oil & CO1& Futures, ICE\\
Ethanol & ETHNNYPR& Spot, FOB\\
Rapeseed Oil & RPSOCRDU & Futures, EU Mill\\
Soybeans & S1& Futures, CBOT\\
Sugarcane & SB1& Futures, ICE\\
Wheat& W1& Futures, CBOT\\
\midrule
\end{tabular}
\end{table}

We analyze time and frequency dependent correlations (wavelet coherence) between biofuels and related commodities. Since our focus is on biodiesel and ethanol, we include only relevant agricultural commodities, which are used for their production, and only relevant fossil fuels, which are their respective natural substitutes. Our dataset thus contains consumer biodiesel ($BD$), ethanol ($E$), corn ($C$), wheat ($W$), soybeans ($S$), rapeseed oil ($RS$), sugarcane ($SC$), crude oil ($CO$), German diesel ($GD$) and the US gasoline ($USG$). Corn, wheat and sugarcane are the feedstock for ethanol; soybeans and rapeseed oil are the feedstock for biodiesel. As ethanol is mainly the US domain and its natural substitute is gasoline, we include the US gasoline. In a similar way, biodiesel is predominantly the EU domain and its substitute is diesel, thence German (as the biggest EU economy) diesel is included. Crude oil (Brent) is included as well because it serves as a production factor for all fuels in our dataset, or at least indirectly. This basic structure of the biofuels system has been validated by our previous analysis in \cite{Kristoufek2012a}. Majority of the dataset was obtained from the Bloomberg database (Table \ref{tab1}), rapeseed oil from the DataStream database, and the two fossil fuels were obtained from the U.S. Energy Information Administration and represent the countries' average price. As the price series of the biofuels are very illiquid, we analyze weekly data for a period between 24.11.2003 and 28.2.2011 (Monday closing prices).

\begin{figure}[htbp]
\center
\begin{tabular}{cc}
\includegraphics[width=2.5in]{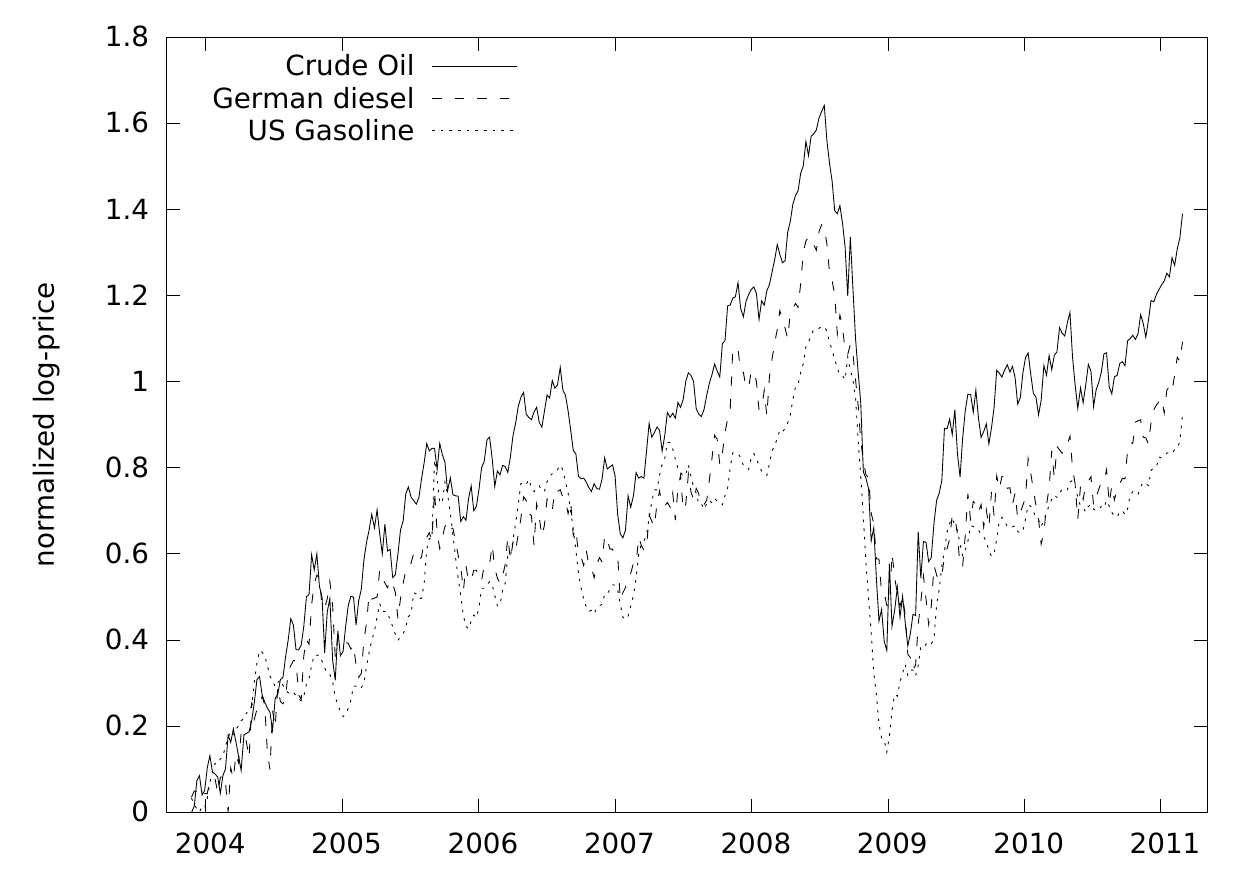}&\includegraphics[width=2.5in]{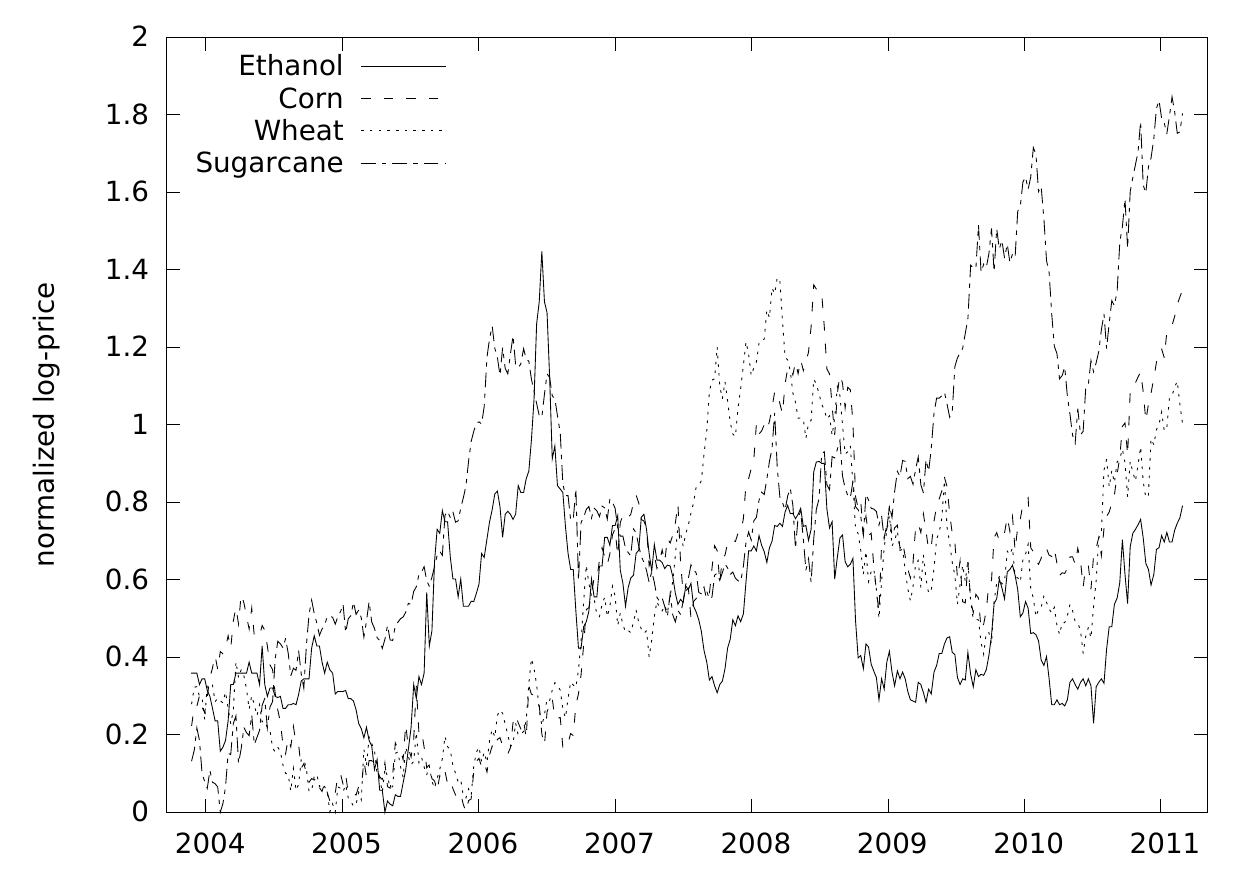}\\
\end{tabular}
\begin{tabular}{c}
\includegraphics[width=2.5in]{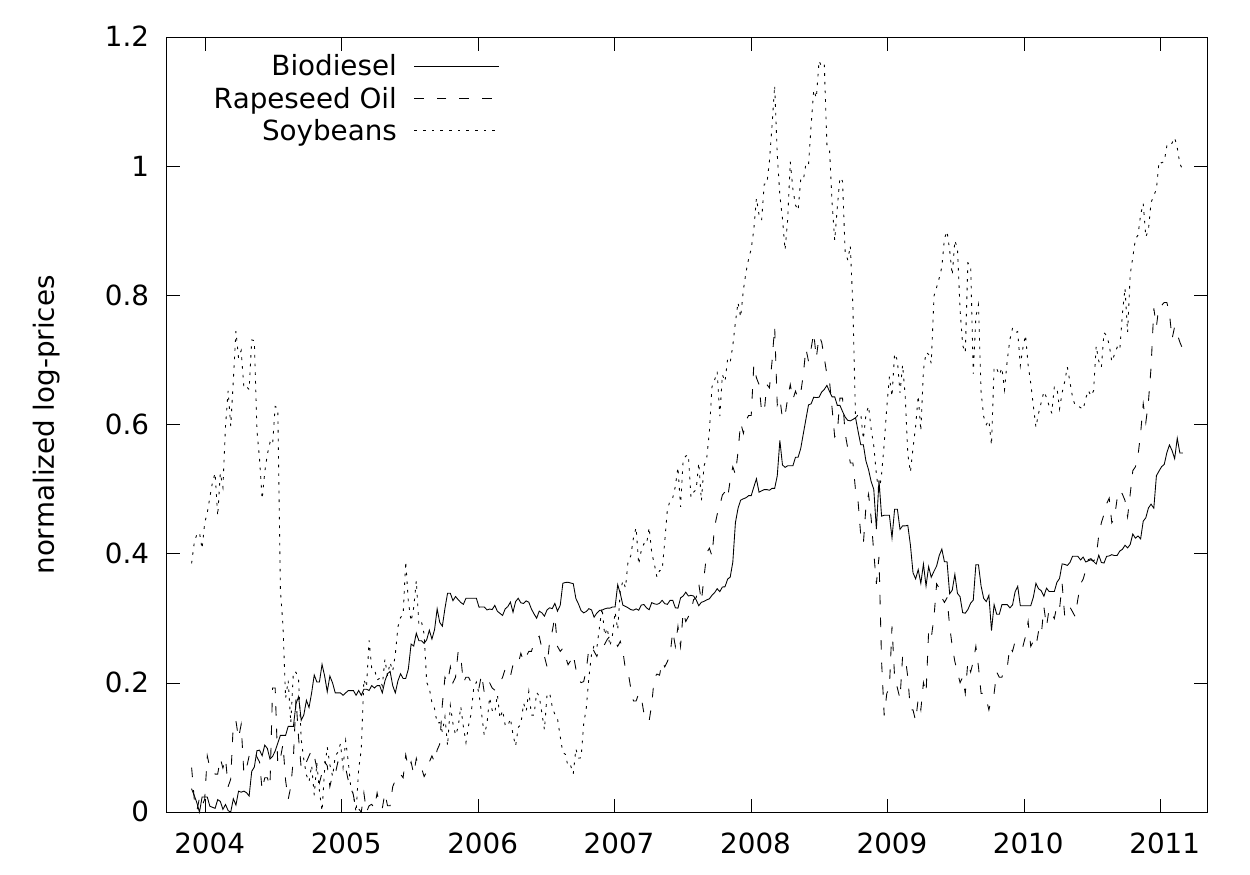}\\
\end{tabular}
\caption{\footnotesize\textit{Logarithmic prices.} Logarithmic prices are normalized so that the minimum value is subtracted making the series more easily comparable.\label{Prices}}
\end{figure}

Fig. \ref{Prices} shows weekly logarithmic prices for all analyzed commodities. The retail fossil fuels are obviously highly correlated with crude oil and the normalized prices almost overlap. Strong increasing trend in prices is observed for the period between 2007 and a middle of 2008 which corresponds to the food crisis period \citep{Hochman_Rajagopal_Timilsina_and_Zilberman_2011}. For the ethanol and related agricultural commodities, the highest prices are connected to the half of 2008. Corn and wheat even reach their maxima in this period. Even though ethanol experienced increasing prices in the food crisis period, these prices are only mildly higher than the heights of 2007 and are even much lower than the maximum in 2006. Sugarcane seems rather unconnected with the rest of the ethanol feedstock commodities and shows the highest variability while reaching its maximum at the break of 2010 and 2011. For the biodiesel and its feedstock commodities, we observe that biodiesel itself has a relatively stable price with slow increasing trend between 2004 and the end of 2005 followed by the period between 2006 and the second half of 2007, where the prices remained very stable. During the food crisis, the price of biodiesel rocketed reaching the peak in the middle of 2008 and returning to the previous levels the following year. Rapeseed oil follows relatively similar path to biodiesel but is much more volatile while soybeans are even more variable in time. Again, the period of food crisis is connected to strong local maxima of the three commodities.

Out of all 45 possible pairs of commodities in our dataset, we focused only on two biofuels branches -- the ethanol (ethanol, corn, wheat, sugarcane, crude oil and the US gasoline) and biodiesel branch (biodiesel, soybeans, rapeseed oil, crude oil and German diesel) -- and analyzed only the relevant connections as a follow-up to our previous results \citep{Kristoufek2012a}. As we are primarily interested only in pairs including a biofuel, we were left with 9 pairs to analyze. Wavelet coherence analysis is applied on the logarithmic returns of weekly prices.

\begin{figure}[htbp]
\center
\begin{tabular}{cc}
\includegraphics[width=3in]{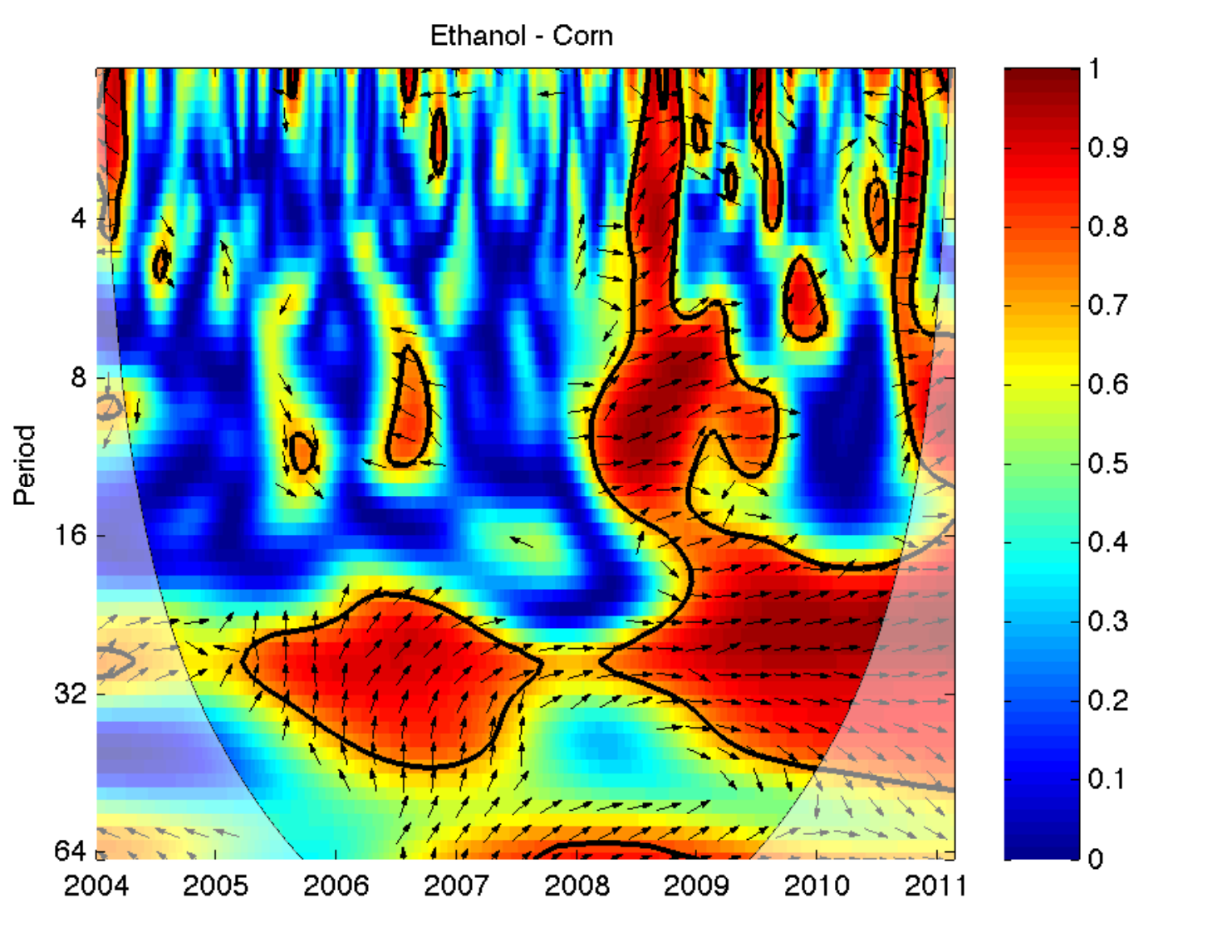}&\includegraphics[width=3in]{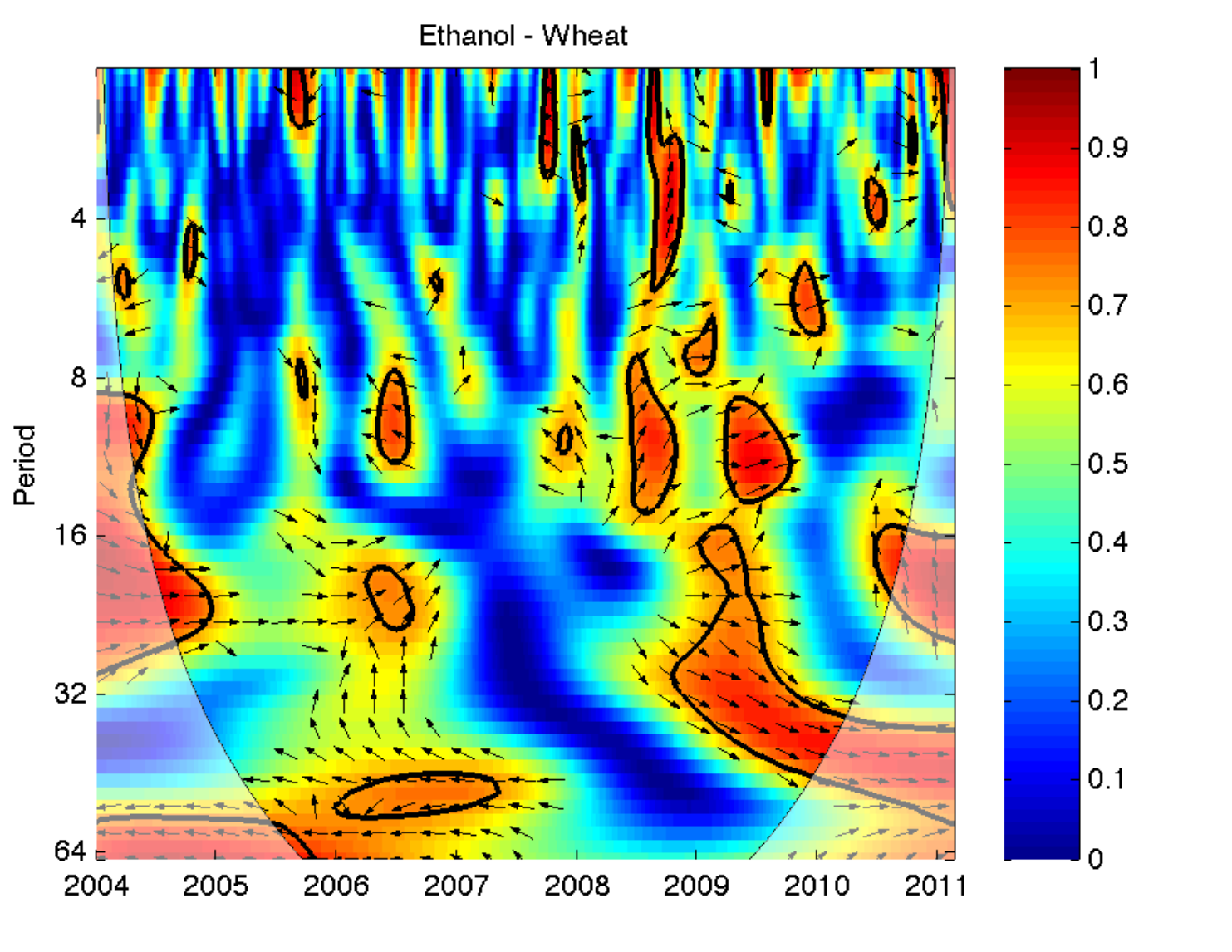}\\
\includegraphics[width=3in]{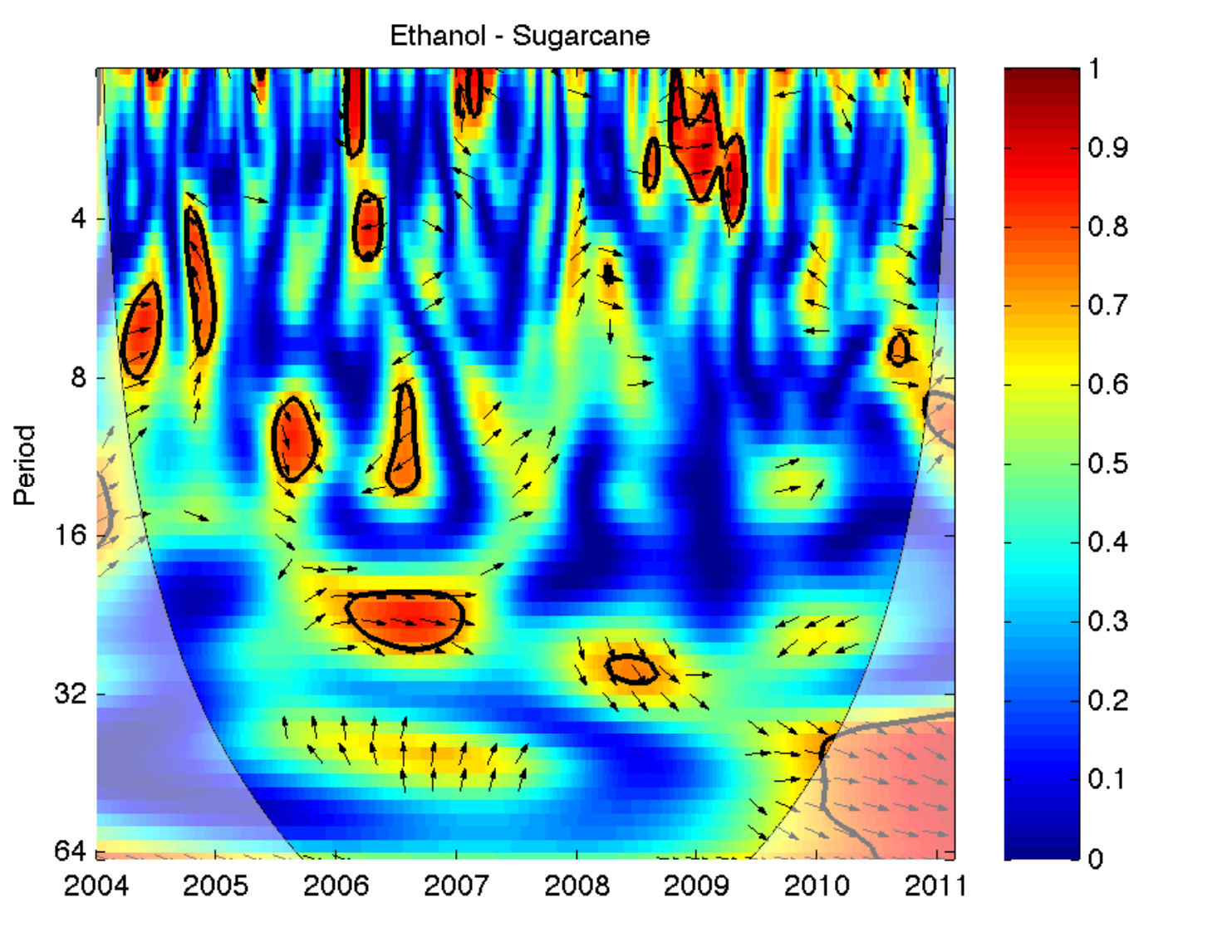}&\includegraphics[width=3in]{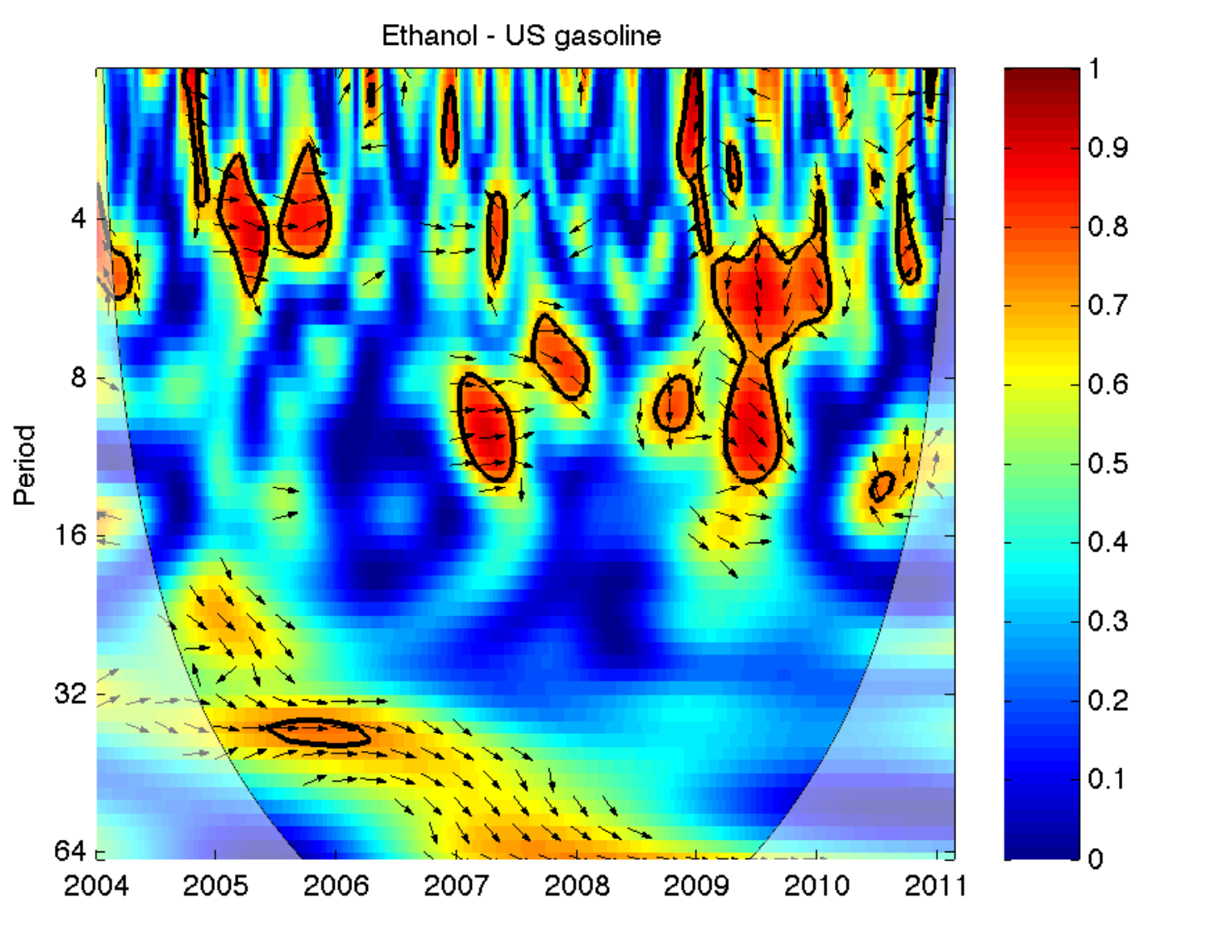}\\
\end{tabular}
\begin{tabular}{c}
\includegraphics[width=3in]{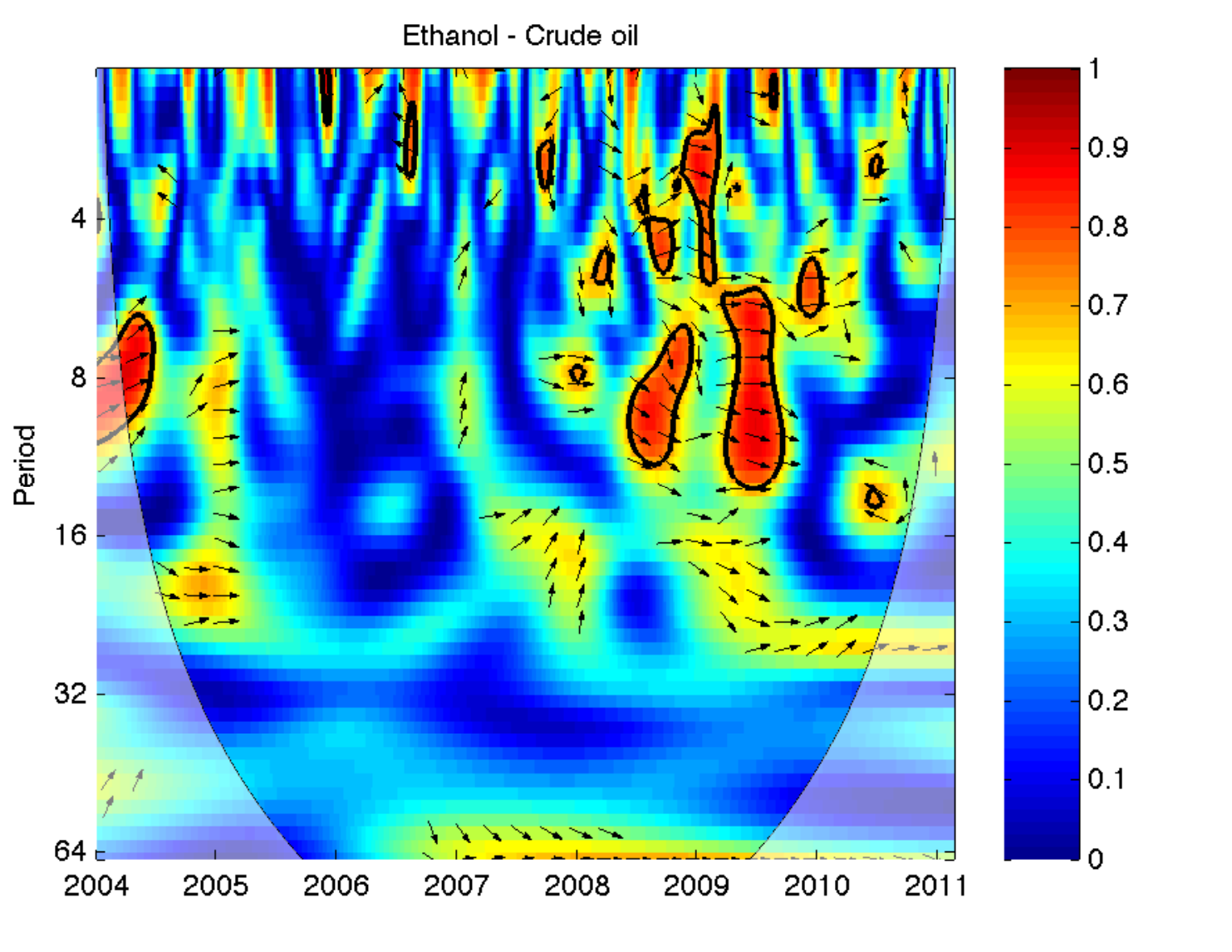}\\
\end{tabular}
\caption{\footnotesize\textit{Wavelet coherence for ethanol and related commodities.}\label{E-C}}
\end{figure}

Starting with the ethanol branch, we found that out of five analyzed pairs, only the ethanol -- corn pair shows economically interesting and statistically significant results. In Fig. \ref{E-C}, we present wavelet coherence for the ethanol branch. There are several features needing further description -- the wavelet coherence can be seen as correlation between analyzed commodities and here, the hotter the color, the higher the correlation; regions of statistically significant correlations are bordered with a bold black line (against the null hypothesis of red noise, i.e. AR(1)-noise); and the direction of correlations is marked by an arrow as described in the previous section. From the picture, we can tell than in the first half of the analyzed sample, ethanol and corn are only weakly correlated and this statistically significant correlation occurs only for scales approximately between a quarter and one year in the period between half of 2005 and half of 2007. In this period, corn clearly leads ethanol. Note that when the arrow points straight upwards, then corn leads ethanol by $\frac{\pi}{2}$, 
i.e. by one quarter of the corresponding scale. With this in mind, we can say that corn leads ethanol by approximately two months in the second half of 2005 while the leading period, i.e. a lag between the two series, was decreasing from 2006 onwards and even reached insignificant correlations at all scales at the break of 2007 and 2008. Starting from 2008, we observe a rapid increase of correlations at almost all scales. Note that this period is connected to very high prices of all the analyzed commodities -- the food crisis. For lower frequencies (higher scales), we observe that corn and ethanol are highly correlated but it is not clear which commodity is the leading one. The higher the frequency gets, the more visible it becomes that corn leads ethanol. Moving forward in time, we can see that from the beginning of 2009 onwards, the dominating frequencies lower considerably and the relationship becomes the most evident approximately between one and three quarters of the year. However, compared to the relationships before the food crisis, we find no dominance between the two. For the remaining pairs, i.e. $E$--$W$, $E$--$CO$, $E$--$SC$ and $E$--$USG$, we find no economically interesting and significant relations between the series and if so, these are rather short-term and can be hardly distinguished from random occurrences as shown in Figs. \ref{excoher}-\ref{exphase}. 

\begin{figure}[htbp]
\center
\begin{tabular}{cc}
\includegraphics[width=3in]{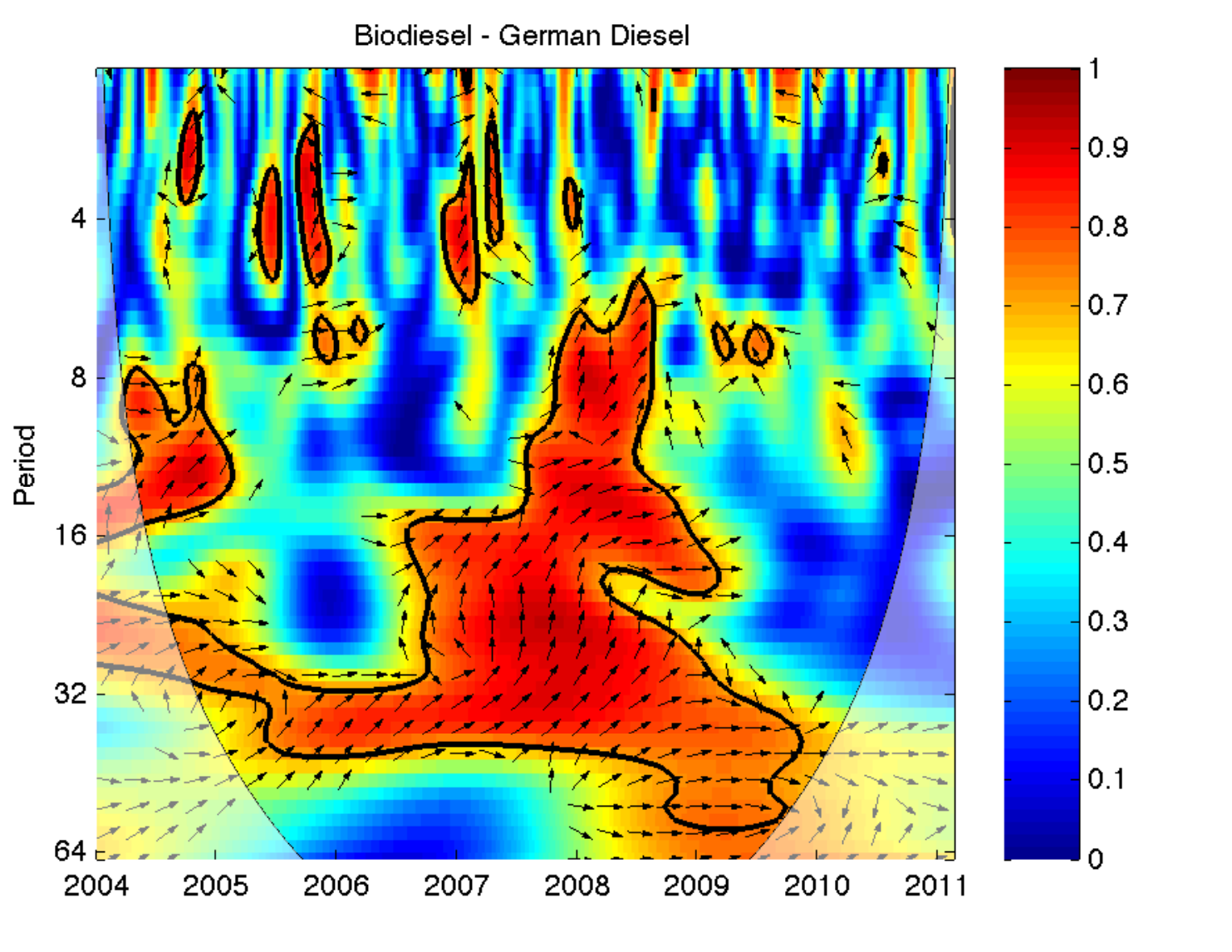}&\includegraphics[width=3in]{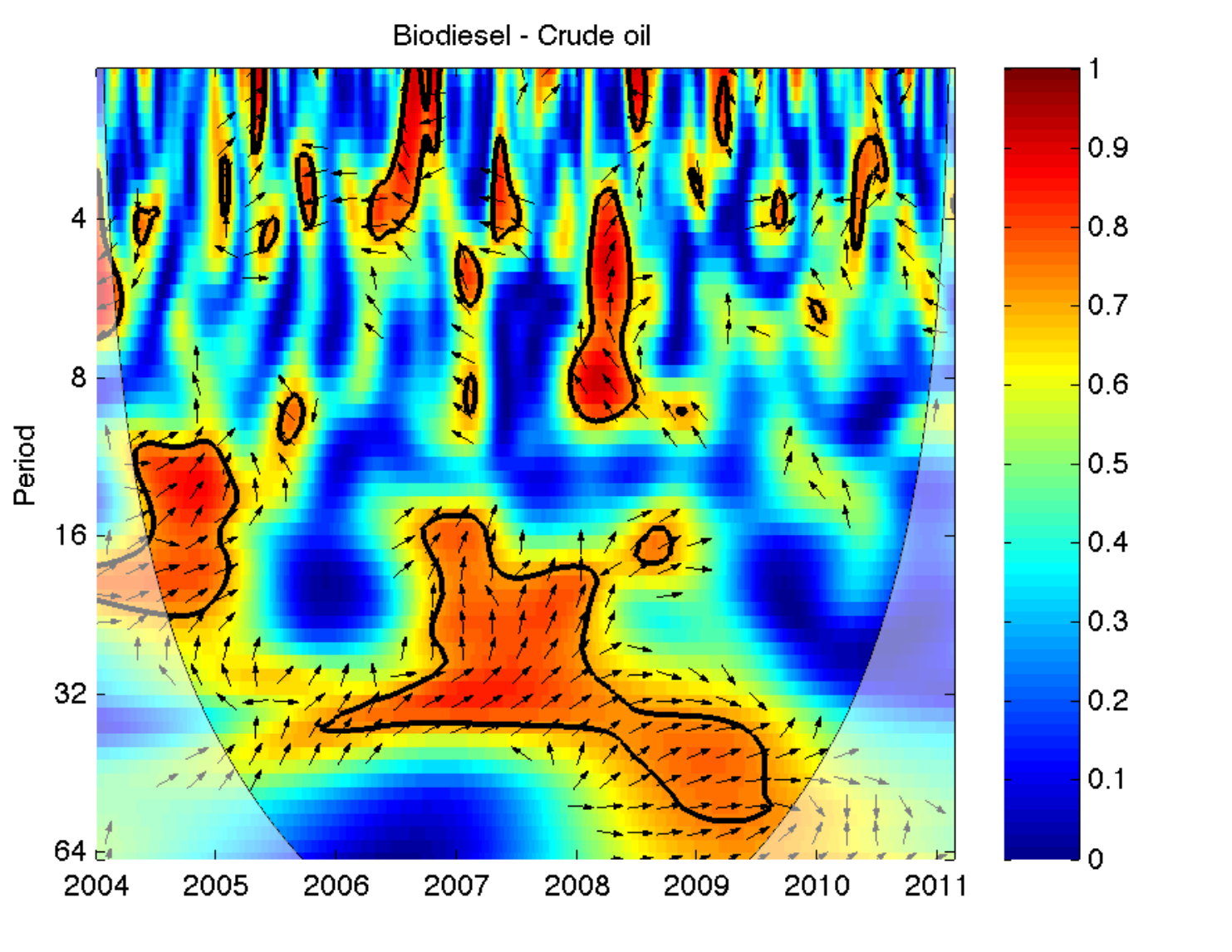}\\
\includegraphics[width=3in]{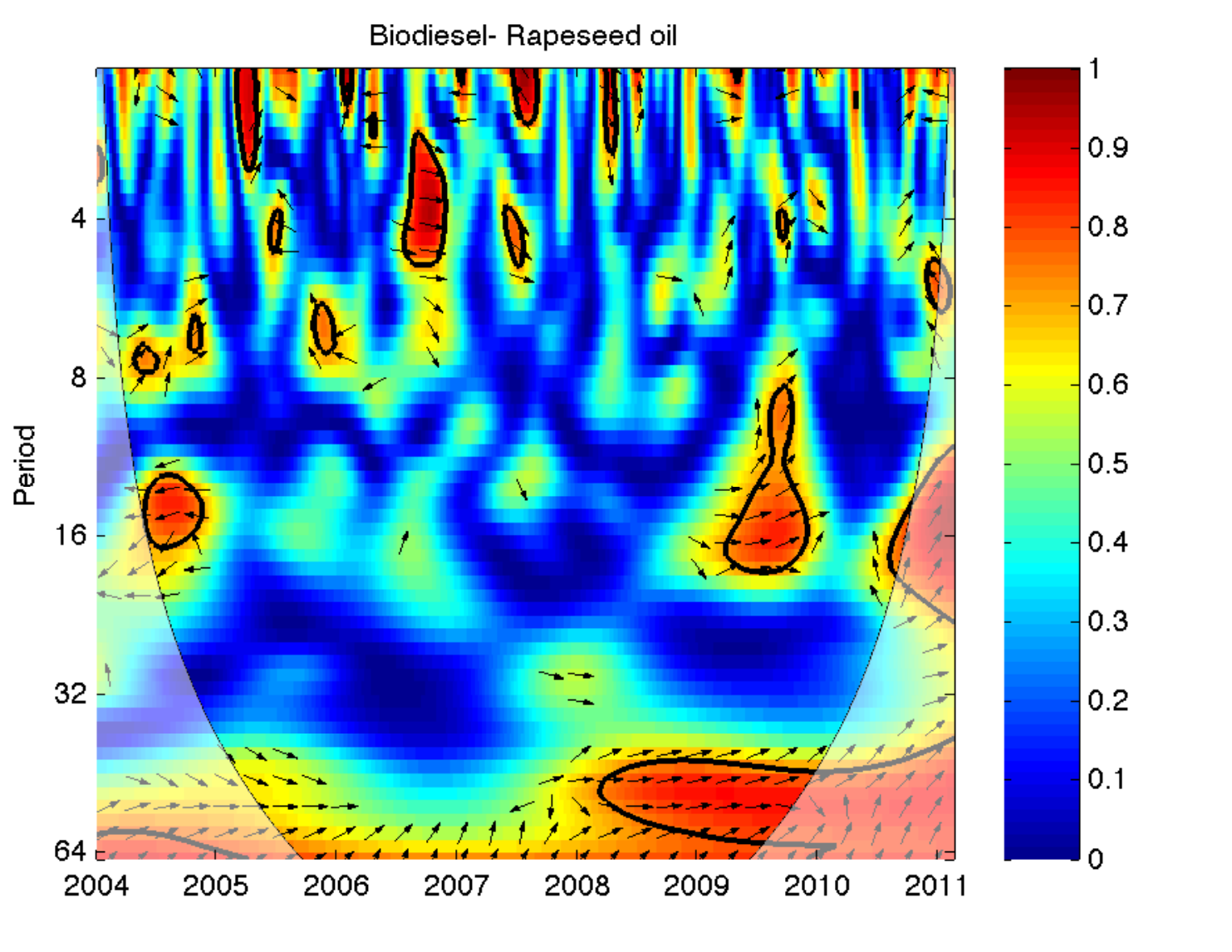}&\includegraphics[width=3in]{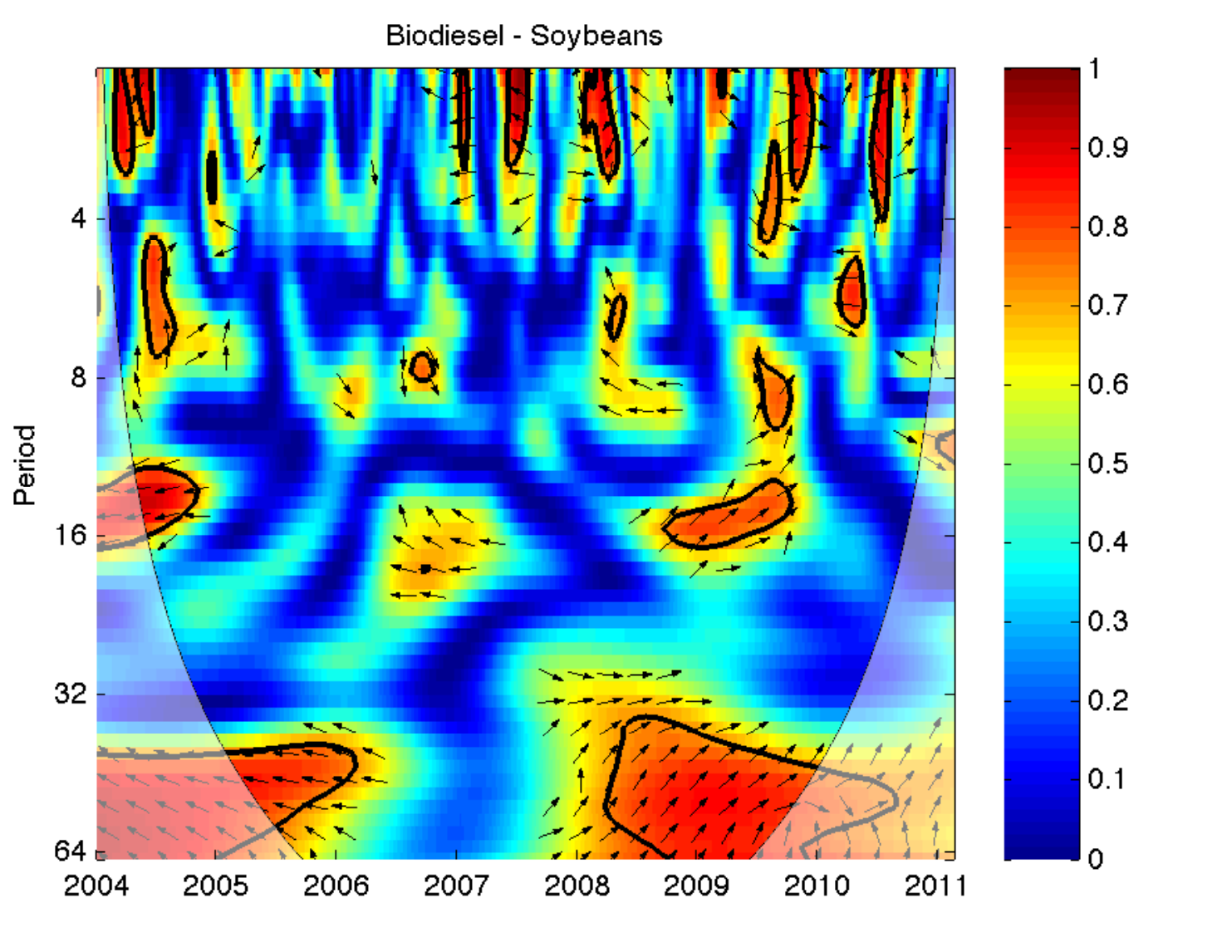}\\

\end{tabular}
\caption{\footnotesize\textit{Wavelet coherence for biodiesel and related commodities.}\label{BD-GD}}
\end{figure}

Moving to the biodiesel branch, we find that the pair with the most pronounced interactions is the biodiesel and German diesel one. In Fig. \ref{BD-GD}, we can see that the most dominant scale is approximately 32 weeks for almost whole analyzed period. Biodiesel and German diesel are positively correlated and in majority of cases, German diesel is the leading series. However, the length of the lag between commodities is on average shorter than for $E$--$C$ case, i.e. biodiesel reacts faster to changes in German diesel than ethanol does to the changes in corn. In the beginning of 2007, German diesel started a growth rally which culminated in a half of 2008. This period is connected with more complex dynamics of correlations between $GD$ and $BD$ with scales of significant correlations broadening to a range between 5 and 50 weeks. German diesel remains the leader of biodiesel for practically all significant scales in this period. For high frequencies between 5 and 10 weeks, we observe a strong leadership of German diesel where the leading period length gets as low as 1--2 weeks. This implies that when the price of German diesel is very high, biodiesel reacts to its changes very quickly. When prices of German diesel get back to the pre-crisis levels -- from the beginning of 2009 onwards -- the dominance of longer scales becomes apparent again. Similarly to the ethanol--corn case, when we compare the pre-crisis and post-crisis correlations at the low frequencies, we have German diesel as a clear leader in the former but no obvious leadership in the latter period. Quite similar, yet much weaker connections are observed for biodiesel and crude oil pair. However, significant connections are visible for only very specific time periods and compared to the $BD$--$GD$ and $E$--$C$ pairs, the coherence is much less evident. Nevertheless, crude oil is identified as a leading series of biodiesel for these significant periods and the series are positively correlated most of the time. The remaining pairs, i.e. $BD$--$S$ and $BD$--$RO$, show practically no significant co-movements.

\section{Conclusions and discussion}

We analyzed the interconnections in ethanol and biodiesel systems with a use of the wavelet coherence analysis, which has never been done before. By doing so, we were able to uncover how correlations between pairs of commodities evolve in time and across frequencies. This way, we overcame the basic problem of standardly used methodologies, i.e. focusing on either the time or frequency domain. Moreover, we did not have to impose any restrictions on the underlying processes as the used methodology is model-free.

Starting with a wide range of the biofuels-related commodities, and covering the most important producing factors and the fossil fuel substitutes for ethanol and biodiesel, we find that the only economically important and statistically significant connections come up between ethanol and corn, and German diesel and biodiesel. For both pairs, we find that the most dominant frequency is around 32 weeks, i.e. approximately half of a year, which holds for almost the whole analyzed period 2003--2011. We also find that a structure of correlations changes with respect to the food crisis between 2007 and 2008, which was connected to unprecedentedly high prices of almost all biofuels feedstock commodities. During this period, the strong interactions in the pairs broadened to higher frequencies as well and the leadership of the producing factors (corn and German diesel) became more apparent. In the crisis period, the leadership of corn relative to ethanol is apparent only for the short scales whereas the German diesel leadership with respect to biodiesel is visible at practically all significant scales.

Interesting distinction between the two pairs of commodities lies in the difference in leadership shifts before and after the food crisis. For ethanol, we observe that corn evidently leads the biofuel at lower frequencies for the pre-crisis period but after the crisis, we find no such strong leadership but only a strong positive correlation between the two. The structure of correlations thus visibly changed after the crisis. Quite similarly, the leadership of German diesel with respect to biodiesel differs before and after the crisis -- the phase shift between the two becomes weaker in time at low frequencies. However, the change is not as noticeable as for the ethanol--corn pair. 

Importantly, we find no evidence for potential squeeze-out effect of agricultural commodities by biofuels, which is of high economic, political and also social interest. On contrary, we find that if some leader-follower relationship is found, the producing factor (corn and German diesel) is the leader of the biofuel (ethanol and biodiesel) in a majority of the cases (both in time and across frequencies), and not vice versa.
   
Note that results presented here nicely integrate and validate partial results of our previous research in \cite{Kristoufek2012a} and \cite{Kristoufek2012}. In \cite{Kristoufek2012a}, we show that from the whole period 2003--2011 viewpoint, there are hardly any correlations between biofuels and the rest of the system at weekly frequency, which changes when we decrease the frequency to one month so that the correlations increase. Importantly, we show that correlations are much stronger for the crisis and post-crisis periods even for high frequencies. This is practically the same result we find with the wavelet coherence analysis. In \cite{Kristoufek2012}, we find that ethanol is lead by corn and biodiesel is lead by German diesel with a use of Granger causality tests, while other connections remain very weak. Again, this is what we show in this paper where we integrate separate correlation and causality techniques, which we used in the earlier papers, by wavelet coherence methodology. Summarizing the results together, we arrive at the very convincing evidence that ethanol (biodiesel) is mainly connected to and lead by corn (German diesel) while the intensity of leadership and magnitude of correlation vary in time and seem to be dependent on corn (German diesel) prices.

Our results show that the wavelet coherence technique is an exceptionally promising technique for analyzing not only biofuels but also the time and frequency dynamics of other commodities. While we introduce this technique in the price domain, it could be obviously used for equally interesting biofuels quantities related analysis as soon as the biofuels markets trading reach such maturity that sufficiently frequent quantity data would be available.

\section*{Acknowledgments}
The authors acknowledge the support from Energy Biosciences Institute at University of California, Berkeley. Karel Janda acknowledges research support provided during his long-term visits at University of California, Berkeley and Australian National University (EUOSSIC program). The support from the Grant Agency of Charles University (GAUK) under projects 118310 and SVV 265 504, Grant Agency of the Czech Republic (GACR) under projects P402/11/0948 and 402/09/0965, and from institutional support grant VSE IP100040 are gratefully acknowledged.We thank Aslak Grinsted for providing us with the MATLAB wavelet coherence package. We thank participants at Bioeconomy Conference in Berkeley, conference of European Association of Environmental and Resource Economists, Australasian Meeting of Econometric Society and seminar participants at Charles University and Australian National University for their valuable comments on earlier versions of this paper. The views expressed here are those of the authors and not necessarily those of our institutions. All remaining errors are solely our responsibility.
 
\section*{References}
\bibliography{BioWavelets}
\bibliographystyle{chicago}

\end{document}